\documentclass[runningheads]{llncs}
\usepackage[T1]{fontenc}

\usepackage{graphicx}

\usepackage{qcircuit}
\usepackage{subcaption}
\usepackage{amsmath}
\usepackage[dvipsnames]{xcolor}

\begin{document}
\title{Towards Federated Learning on the Quantum Internet}

\author{Leo Sünkel \and
Michael Kölle  \and
Tobias Rohe  \and
Thomas Gabor }
\authorrunning{L. Sünkel et al.}

\institute{Institute for Informatics, LMU Munich, Germany \\
\email{leo.suenkel@ifi.lmu.de}
}

\maketitle           
\begin{abstract}
While the majority of focus in quantum computing has so far been on monolithic quantum systems, quantum communication networks and the quantum internet in particular are increasingly receiving attention from researchers and industry alike. The quantum internet may allow a plethora of applications such as distributed or blind quantum computing, though research still is at an early stage, both for its physical implementation as well as algorithms; thus suitable applications are an open research question. We evaluate a potential application for the quantum internet, namely quantum federated learning. 
We run experiments under different settings in various scenarios (e.g. network constraints) using several datasets from different domains and show that (1) quantum federated learning is a valid alternative for regular training and (2) network topology and nature of training are crucial considerations as they may drastically influence the models performance. The results indicate that more comprehensive research is required to optimally deploy quantum federated learning on a potential quantum internet.

\end{abstract}
\section{Introduction}\label{sec:introduction}
Establishing a quantum communication network over large distances and thereby connecting a multitude of quantum devices with varying architectures and capabilities may allow the \textit{quantum internet} to rise. However, what exactly such a potential network should look like remains an active research question \cite{caleffi2018quantum,simon2017towards,van2022quantum,wehner2018quantum}.
Given that the field is still largely in its infancy, it is vital to identify and examine the potential applications that such a large-scale network could enable. Distributed quantum computing \cite{caleffi2022distributed}, quantum key distribution, blind quantum computing, and quantum federated learning \cite{chen2021federated,wang2023quantum} are all applications that have been discussed in recent years, and more are yet to be discovered. As quantum machine learning has sparked an interest in a wide range of disciplines in recent years, it is no surprise that, sooner or later, combining this area with the field of quantum communication will increasingly attract attention as this provides new opportunities and research avenues. Quantum federated learning already is a step in this direction, and it is this field that is the main topic of discussion in this paper. The idea behind federated learning is to train a global model by a collection of clients that communicate over a network. The crucial point is that each client trains their model on their local dataset, i.e., they keep their data private. Rather than exchanging private data, clients only communicate their models weights; the global model is trained by aggregating weights from participating clients. This (classical) approach can easily be transferred to the quantum domain, resulting in quantum federated learning. However, this field is also still in its infancy and thus many open research questions remain. In this paper, we evaluate different approaches to quantum federated learning under varies constraints that may be present in a quantum internet. More specifically, we run experiments using two different network topologies where the number of qubits of the quantum clients differ. Furthermore, we run experiments where clients are trained on subsets of the same dataset as well where each client is trained on a distinct one. We show that quantum federated learning is a compelling alternative to regular model training while it is crucial to take certain network constraints (e.g. each nodes qubit capacity) as well as the models training approach (e.g. nature of weight aggregation and exchange) into account. This paper is structured as follows. In Section \ref{sec:background} we cover the background of quantum communication networks, variational quantum circuits as well as quantum federated learning and discuss related work in Section \ref{sec:related_work}. We illustrate our approach and architecture in Section \ref{sec:approach} and discuss our experimental setup in Section \ref{sec:experimental_setup}. We present our results in Section \ref{sec:results} and conclude in Section \ref{sec:conclusion}.

\section{Background}\label{sec:background}
We begin this section with a brief recapitulation of the basic building blocks of quantum communication networks (QCNs) and what we understand as the quantum internet. We then discuss potential applications and what this novel form of communication means for quantum algorithms in general while also for quantum machine learning in particular. However, we will not be covering the basics of quantum computation and information (e.g. qubits, entanglement, superposition) and instead refer the reader to other resources \cite{nielsen2010quantum,watrous2018theory} for an in depth introduction to these topics. Once we have reviewed the fundamental concepts of quantum networks, we will discuss federated quantum machine learning, the main topic of this paper.

\subsection{Quantum Networks and the Quantum Internet}
\begin{figure}
    \centering
    \includegraphics[scale=0.4]{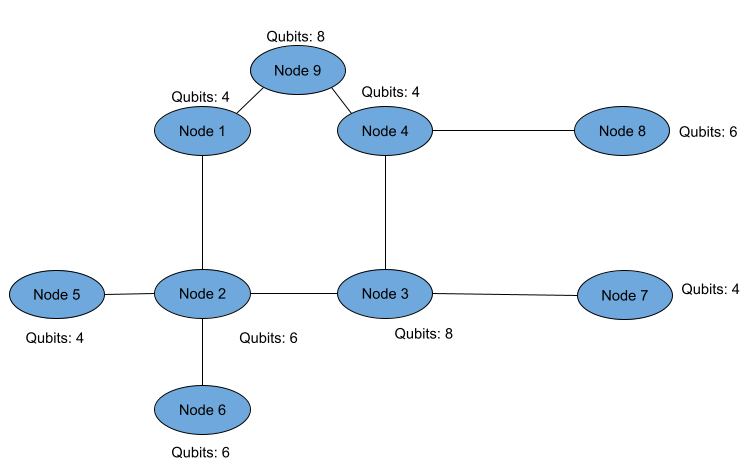}
    \caption{Example of a network with random topology where nodes represent available QPUs. The QPUs in this network have varying capacity, i.e., number of qubits and the network is not fully connected. Both facts cause ramifications and constraints for applications and network design.  }
    \label{fig:example_network}
\end{figure}

We define a QCN as a number of quantum computers that are connected via classical and quantum channels, i.e., they have the means to communicate classical messages as well as qubits or quantum states with each other. We will refer to these quantum computers as QPUs or nodes in the network, and we use these terms interchangeably. Furthermore, a QCN can be seen as a graph where the nodes are quantum devices (QPUs) and the edges are the communication channels. An example of such a simple network can be seen in Figure \ref{fig:example_network}. This example network consists of 9 nodes (QPUs) each with a different qubit capacity. Each edge in the network serves a dual purpose, facilitating both classical and quantum communications.

Unfortunately, quantum channels can be noisy, thus information loss is inevitable. Moreover, quantum information cannot be copied as this is prohibited by the \textit{no cloning theorem}, and so classical protocols to overcome information loss as well as sending messages cannot be transferred into the quantum realm in a straightforward manner. In a QCN, qubits are \textit{teleported} over quantum channels to other nodes in the network. However, qubits are not physically teleported, it is the state that is transferred to another qubit, however, the state of the original qubit is destroyed in this process, that is, the qubit cannot be copied. The teleportation protocol requires an \textit{EPR-pair}, i.e., two qubits that are maximally entangled as well as a classical and a quantum channel. In order to teleport qubits over large distances, \textit{quantum repeaters} can be deployed along the way to combat decoherence, i.e., the loss of information. Quantum repeaters perform \textit{entanglement swapping} and possibly \textit{entanglement distillation} that (1) transfer the quantum state and (2) increase the entanglement fidelity. 

With these building blocks established, one can envision a large quantum network that is akin to a so-called quantum internet \cite{kimble2008quantum}. However, how concretely such a quantum internet exactly should look like is up to debate and a crucial research question in quantum communication and quantum computing alike \cite{caleffi2018quantum,kozlowski2019towards,simon2017towards,van2022quantum,wehner2018quantum}. A quantum internet may connect QPUs with vastly different architectures, i.e., consist of heterogeneous nodes, a fact that will become relevant later in this work. 

In summary, a quantum network consists of multiple QPUs connected via classical and quantum channels that allow them to exchange classical as well as quantum information. However, due to various constraints imposed by the laws of quantum physics, many of the classical protocols cannot be transferred to the quantum setting. Moreover, decoherence and the loss of information are major challenges and while countermeasures have been proposed, more research is required to establish a working quantum internet. 

This introduction is intended as a short recap of the fundamentals required to follow the work performed in this paper, for an in depth introduction we refer to \cite{caleffi2022distributed,illiano2022quantum,van2013designing}.

\subsection{Variational Quantum Circuits}
Variational Quantum Circuits (VQCs)\cite{cerezo2021variational} stand out from classical circuits by leveraging quantum phenomena such as superposition and entanglement. These circuits consist of parameterized gates that are optimized through classical techniques, making them a hybrid approach suited for tasks like quantum machine learning. We summarize the essential components of VQCs in this section and describe how we apply them in the approach introduced in Section \ref{sec:approach}.

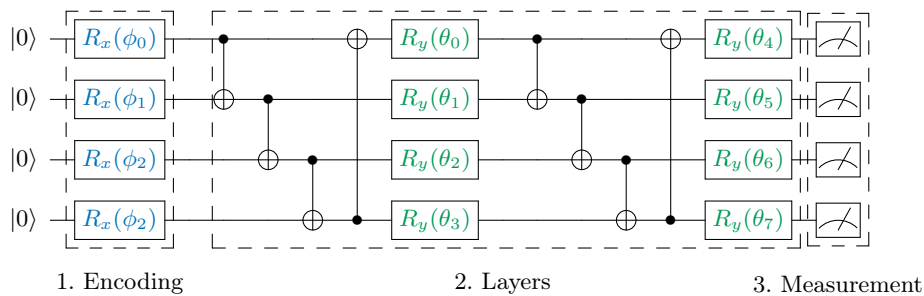
\begin{figure}
    \centering
    \begin{equation*}
        \Qcircuit @C=1em @R=.9em{
           \lstick{|0\rangle} & \gate{\color{RoyalBlue}R_x(\phi_0)} & \qw & \ctrl{1} & \qw & \qw & \targ & \gate{\color{ForestGreen}R_y(\theta_0)} & \qw & \ctrl{1} & \qw & \qw & \targ & \gate{\color{ForestGreen}R_y(\theta_4)} & \meter  \\
           \lstick{|0\rangle} & \gate{\color{RoyalBlue}R_x(\phi_1)} & \qw & \targ & \ctrl{1} & \qw & \qw & \gate{\color{ForestGreen}R_y(\theta_1)}  & \qw & \targ & \ctrl{1} & \qw & \qw & \gate{\color{ForestGreen}R_y(\theta_5)} & \meter \\ 
           \lstick{|0\rangle} & \gate{\color{RoyalBlue}R_x(\phi_2)} & \qw & \qw & \targ & \ctrl{1} & \qw & \gate{\color{ForestGreen}R_y(\theta_2)} & \qw & \qw & \targ & \ctrl{1} & \qw & \gate{\color{ForestGreen}R_y(\theta_6)} & \meter \\
           \lstick{|0\rangle} & \gate{\color{RoyalBlue}R_x(\phi_2)} & \qw & \qw & \qw & \targ & \ctrl{-3} & \gate{\color{ForestGreen}R_y(\theta_3)} & \qw & \qw & \qw & \targ & \ctrl{-3} & \gate{\color{ForestGreen}R_y(\theta_7)} & \meter \gategroup{1}{2}{4}{2}{.7em}{--} \gategroup{1}{4}{4}{14}{.7em}{--} \gategroup{1}{15}{4}{15}{.7em}{--} \\
           & & & & & & & & & & & & & & & \\
           & \mbox{1. Encoding} & & & & & & & \mbox{2. Layers} & & & & & & \mbox{3. Measurement} 
        }
    \end{equation*}
    \caption{Example of a variational quantum circuit (VQC): Features are embedded through rotations (depicted blue) in the first step. This is followed by two repeating layers of CNOT-gates and parameterized rotations (green), the $\theta$ values are the weights being optimized. In the final step, the qubits are measured resulting in a classical output (0 or 1) for each qubit. }
    \label{fig:example_circuit}
\end{figure}

VQCs can, for example, be applied to classification problems \cite{schuld2020circuit} and are a promising hybrid approach to QML in the NISQ-era. As this is a classical-quantum hybrid approach it consists of essentially two parts, namely the quantum circuit and the classical optimization routine. The quantum part, i.e. the VQC, can loosely be divided into three distinct constituents: (i) feature encoding, (ii) a series (layers) of parameterized rotation gates followed by entangling gates and finally (iii) the measurement operations. An example circuit with two layers is depicted in Figure \ref{fig:example_circuit}. The parameters of the rotations correspond to the weights that are optimized by a classical optimizer while the measurement results are interpreted as predictions and are also used by the optimizer. This approach is iterative, i.e., it executes until a number of epochs or steps has been reached. For an in depth introduction to this topic we refer to \cite{cerezo2021variational,mitarai2018quantum,schuld2020circuit}.

\subsection{Quantum Federated Learning}
Before discussing QFL we will first give a short recap of the central idea behind federated learning in general, i.e., its origins in classical machine learning. We will then discuss quantum variants of this learning approach.  

Federated learning (FL) \cite{mcmahan2017communication} allows for the collective training of a global model by multiple clients, each contributing to the model without exposing their private data. This ensures that each client's data remains confidential and local to them. Rather than sharing data, clients transmit their model weights to a central server, where these weights are aggregated, updated, and then redistributed to all clients for further training iterations. Moreover, each client may be trained on different subsets or even entirely different datasets. Furthermore, many variants and approaches to FL have been proposed; the approach here is a basic and straightforward one, as we focus on QFL in this paper. For our purposes, it suffices to define FL as an decentralized learning approach in which a global model is trained by multiple clients without revealing their private data. For more comprehensive introduction to the topic we refer to \cite{geyer2017differentially,konevcny2016federated,wang2020federated,yang2019federated}

QFL follows a similar line, however, many different approaches can also be employed here; though we will be focusing on simplicity. In QFL, the client models can be replaced by VQCs, and weights can be communicated through classical as well as quantum channels, though the former does not necessarily require a quantum network while the latter does. Analog to the classical approach, each client (e.g. VQC) is then trained on its own local dataset. We will discuss the peculiar details arising through the quantum internet in Section \ref{sec:approach}.

\section{Related Work}\label{sec:related_work}
Several different approaches and aspects of QFL have been explored by the research community. For example, privacy aspects are investigated in \cite{li_privacy-preserving_2023} and \cite{rofougaran_federated_2023} while combining QFL with blind quantum computing is discussed by Li et al. in \cite{li_quantum_2021} and Zhang et al. propose a quantum method for parameter aggregation in \cite{zhang_federated_2023}. Challenges of QFL in the context of quantum networks are discussed by Chehimi et al. in \cite{chehimi_foundations_2023}. Wang et al. \cite{wang2023quantum} apply QFL on a binary classification task using a ring topology, i.e., without a central model. Moreover, they use quantum weights in their approach. In \cite{chen2021federated} the authors evaluate a hybrid quantum-classical approach on a binary classification task using images.

\section{Approach}\label{sec:approach}
In this section, we describe our approach to QFL. We discuss quantum clients and the overall architecture of our methods. This includes network constraints as well as how the models train and communicate.

\subsection{Quantum Clients}
In our proposed QFL approach, the client model is a VQC running on a quantum node in a (quantum) communication network. More specifically, we consider the quantum internet in which nodes are quantum computers each with potentially a different qubit capacity. It's important to note that the necessity for a quantum communication channel hinges on the specific methodology of weight exchange in the applied QFL approach. That is, weights can be exchanged classically and thus a simple classical channel connecting quantum nodes suffices. Furthermore, in our scenario each node executes the model of only a single client. Thus, the clients in our QFL approach execute VQCs with a different number of qubits and hence number of trainable parameters. Consider the example network in Figure \ref{fig:example_network}. Here, 9 nodes are connected to form a network where nodes have different qubit capacities.

\subsection{Architecture}
\begin{figure}[t]
    \centering
    \includegraphics[scale=0.3]{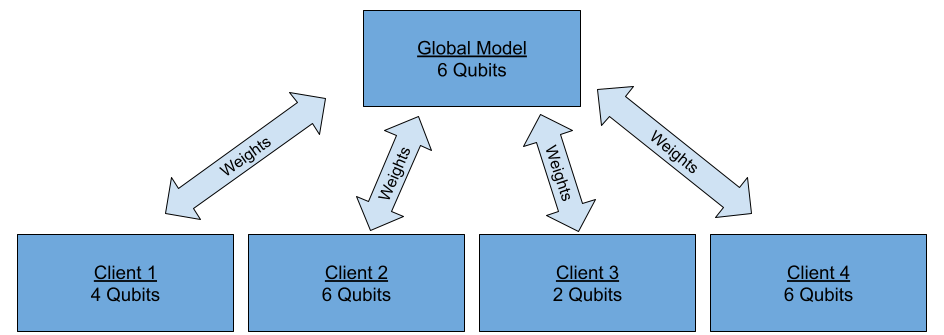}
    \caption{Example of a QFL approach where a global model is collectively trained by 4 different clients with different qubit capacities. Clients send their weights to the global model that aggregates, updates and distributes the new weights for the next round of training.}
    \label{fig:example_global_model_topology}  
\end{figure}

We evaluate two different approaches to QFL as part of this work that differ in the topology of the arrangement of clients and how weights are aggregated and exchanged. In the first approach, a global model is collectively trained by multiple clients who only exchange their weights with the global model. The global model is then responsible for aggregating, updating and distributing the weights among all participants in each training round. An example of this architecture is depicted in Figure \ref{fig:example_global_model_topology}.

\begin{figure}[bt]
    \centering
    \includegraphics[scale=0.3]{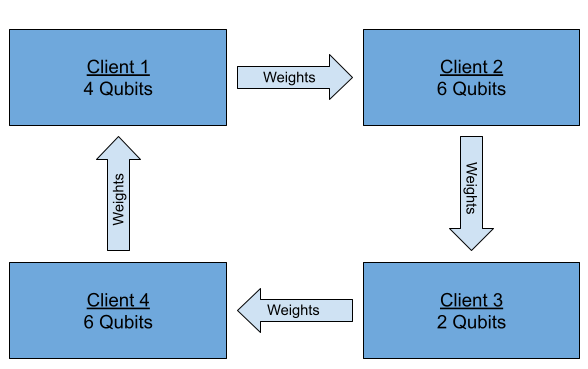}
    \caption{Example of clients with varying qubit capacity arranged in a ring topology. In this scenario, when clients transfer their weights to the next, the number of weights is adjusted accordingly.}
    \label{fig:example_ring_topology}
\end{figure}

In the second approach, clients are arranged in a ring topology (cf. \cite{wang2023quantum}) and no global model is trained. More specifically, clients train their model using their local dataset and send their weights to the  succeeding client after completing a training round. This approach is shown in Figure \ref{fig:example_ring_topology}. As the clients capacity, i.e., number of qubits may vary from one to the next, the number of weights must be adjusted when sent to the following client. In our experiments we adjust the number of weights in the following way. If the next client contains less qubits (and thus weights), the client discards superfluous weights, i.e., it only uses the first \textit{n} weights required. In the case where the next client contains more qubits, the client uses all weights from the previous one while filling up the missing weights with its own from the last round. Note that this is  an ad-hoc approach and the appropriate aggregation of weights in the context of trainability is its own research topic and merits its own discussion, however, is not in the scope of this work.

\subsection{Quantum Circuits}
In our model, we apply two different VQCs that differ only in the embedding method applied and number of qubits. That is, we use angle embedding in one set of experiments and amplitude embedding in another. 

\section{Experimental Setup}\label{sec:experimental_setup}
This section outlines the models, datasets, and parameters used in our experiments.

\subsection{Datasets}
Our experiments utilize three datasets, detailed as follows. Note that we focus solely on binary classification in this work. In datasets that contain more than two classes, we divide the dataset into subsets such that each only contains images of two classes.

\begin{figure}
    \centering
    \includegraphics[scale=0.6]{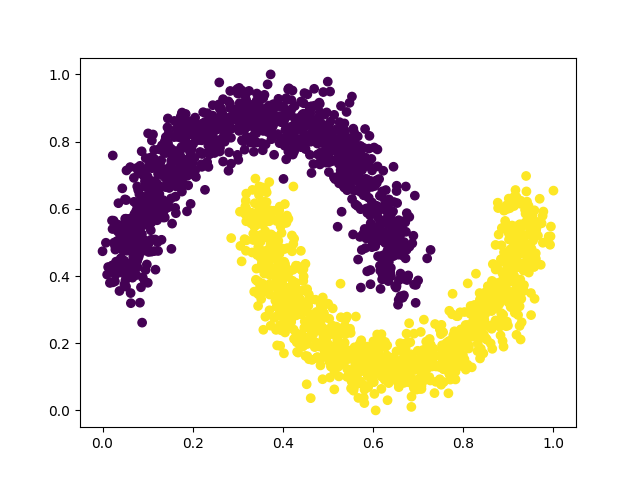}
    \caption{Two half circles ("moons" dataset) with added noise factor of 0.1 created with sci-kit learn.}
    \label{fig:moons_data}
\end{figure}

\textbf{Moons:} The moons dataset contains 2 features for a binary classification problem where each class is shaped as a half circle and is provided by scikit-learn \cite{scikit-learn}. We used a dataset with 3000 samples. The training set used is visualised in Figure \ref{fig:moons_data}. 

\textbf{FashionMNIST:} The FashionMNIST \cite{xiao2017} dataset contains 70000 grayscale images for 10 different classes where each image has a size of $28\times28$. Note that there are 60.000 training and 10000 test images. We used this dataset provided by PyTorch \cite{paszke2019pytorch}. 

\textbf{PneumoniaMNIST:} The PneumoniaMNIST dataset contains 5856 chest X-ray images with a size of $28\times28$ for binary classification. We used the dataset provided by \cite{medmnistv2}. Note that for all datasets, we used 1000 images per epoch in training.

\textbf{Preprocessing:} Images were resized for QPUs with lesser capacity than required to embed the entire image, the number of features used is shown in Table \ref{tab:parameters}.

\subsection{Training}
In this section, we discuss the two training approaches we apply in our experiments. In the first, multiple clients are trained on a single dataset while in the second each client is trained on a different dataset.

\subsubsection{Multiple clients, single dataset:}
In this experiment, each client is trained on a single dataset (moons), however, the dataset is subdivided into distinct subsets prior, i.e., each client is trained on a different subset of the same dataset. Furthermore, in this experiment each client has an equal number of qubits (2) and uses a VQC with angle embedding as its model. The parameters used are listed in Table \ref{tab:parameters}. This experiment was run using both topologies described in Section \ref{sec:approach}

\begin{table}[]
    \centering
    \begin{tabular}{|l|c|c|c|c|}
    \hline
        & \textbf{MCSD} & \textbf{MCMD} & \textbf{Baseline (MCSD)} & \textbf{Baseline (MCMD)} \\ \hline
        \textbf{Features} & 2 & [16, 64, 784, 784] & 2 & 784 \\ \hline
        \textbf{Depth} & 8 & 10 & 8 & 10 \\ \hline
        \textbf{Qubits} & 2 & [4,  6, 10, 10] & 2 & 10 \\ \hline
        \textbf{Clients} & 3 & 4 & - & - \\ \hline
        \textbf{Embedding} & Angle & Amplitude & Angle & Amplitude\\
    \hline
    \end{tabular}
    \caption{Parameters used in the QFL and quantum baseline experiments. (MCSD=multiple clients single dataset, MCMD=multiple clients multiple datasets). Note that in the experiments where all clients have an equal capacity, the capacity is set to the number of qubits required to embed all features (i.e., 2 for the moons and 10 for the image datasets).}
    \label{tab:parameters}
\end{table}

\subsubsection{Multiple clients, multiple datasets:}
In this experiment, multiple datasets are used in training, that is, each client is trained on a different trainingset or subset. More specifically, we used three distinct subsets of FashionMNIST ("Trouser vs. Pullover", "Sandal vs. Sneaker" and "Dress vs. Coat") and the pneumonia dataset, where each client was trained on one of these sets. For example, client 1 is trained on images depicting trousers and pullovers, client 2 on sandals and sneakers while the third client is trained on chest X-ray images and the fourth on images of dresses and coats, each in a binary classification task. Moreover, each client may have a different capacity (i.e. qubits). This means that each client also embed, and thus are trained, using a different number of features of the input image. Parameters of this approach  are depicted in Table \ref{tab:parameters}. Like the first experiment, this one was also run using both topologies described in Section \ref{sec:approach}.

\section{Results}\label{sec:results}
We discuss the results of all experiments conducted as part of this work in this section. We first examine the experiments with multiple clients, single dataset using both QFL approaches described earlier. We then move on to the experiments with multiple clients, multiple datasts. All experiments were run with 5 different seeds and plots depict the mean of the aggregated results. Note that for baselines, we aggregated the results of all individual binary classification experiments, which is a crucial fact to keep in mind when comparing all approaches. Moreover, in the test results of the experiments using the ring topology, each client uses the final parameters after training for its own model on its respective test dataset and the results depicted below are aggregated over all clients.

\subsection{Multiple Clients, Single Dataset}
\begin{figure}[]
    \begin{subfigure}{0.5\textwidth}
    \includegraphics[scale=0.4]{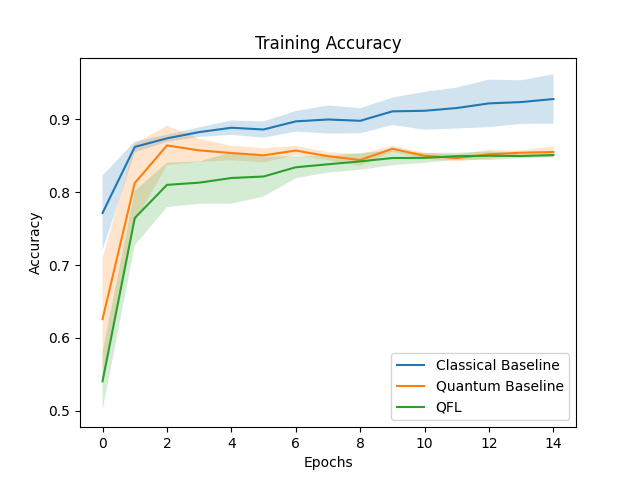}
    \caption{Training accuracy..}
    \label{fig:mcsd_star_train_accuracy}
    \end{subfigure}
    \begin{subfigure}{0.5\textwidth}
    \includegraphics[scale=0.4]{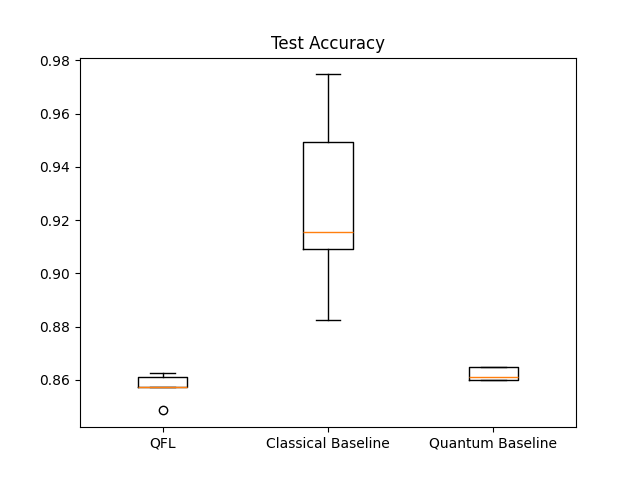}
    \caption{Test accuracy.}
    \label{fig:mcsd_star_test_accuracy}
    \end{subfigure}
    \label{fig:mcsd_star_accuracy}
    \caption{Results for the MCSD experiments using the star topology.}
\end{figure}

Recall that in this setting, we train multiple clients on a single dataset (we abbreviate this approach as MCSD), however, each client receives its own distinct subset. Furthermore, we evaluated this approach on two different QFL architectures, i.e., topologies, namely star and ring. Training and test results for this experiment with the star topology are shown in Figure \ref{fig:mcsd_star_train_accuracy} and Figure \ref{fig:mcsd_star_test_accuracy} respectively. The classical baseline delivers the best training and test accuracy, though this was to be expected as this approach consists of vastly more parameters and a problem of this scale generally is not problematic for classical neural networks. The QFL approach achieves similar performance as the quantum baseline; the difference being only minuscule.

\begin{figure}
    \begin{subfigure}{0.5\textwidth}
        \includegraphics[scale=0.4]{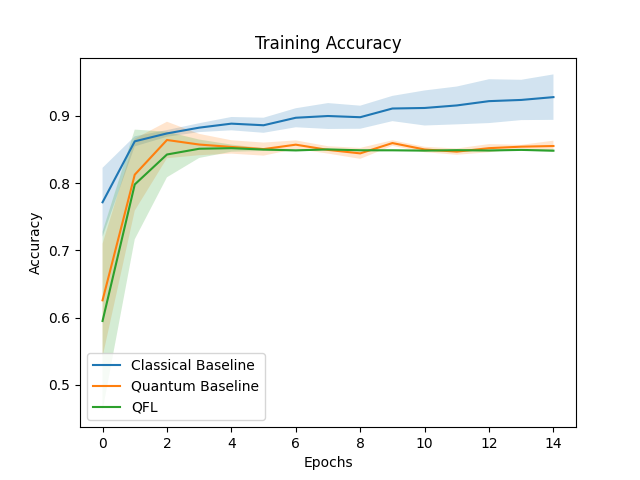}
        \caption{Training  accuracy. }
        \label{fig:mcsd_ring_train_accuracy}
    \end{subfigure}
    \begin{subfigure}{0.5\textwidth}
        \includegraphics[scale=0.4]{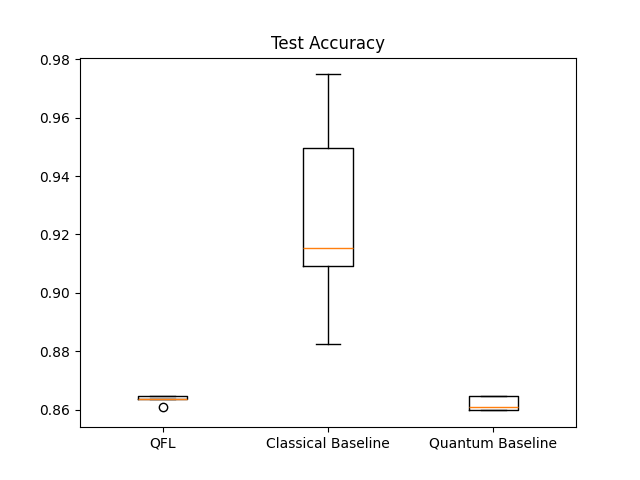}
        \caption{Test accuracy. }
        \label{fig:mcsd_ring_test_accuracy}
    \end{subfigure}
    \caption{Training and test results on MCSD experiments with the ring topology.}
\end{figure}

The train and test results using a ring topology with the moons data is shown in Figure \ref{fig:mcsd_ring_train_accuracy} and \ref{fig:mcsd_ring_test_accuracy} respectively. While the quantum baseline performs slightly better earlier in training, both approaches perform almost identical and converge relatively fast. Testing results are in a similar range for both quantum methods.

\subsection{Multiple Clients, Multiple Datasets}
We discuss the experiments using multiple clients, multiple datasets (MCMD) next. Recall that in these experiments, each client is trained on a different dataset, that is, the first client is trained using images of two distinct classes from FashionMNIST, the second uses two different classes from the same dataset while the third client is trained using the PneumoniaMNIST dataset and a fourth client is trained from two different classes from FashionMNIST. Note that the baselines were also trained on these datasates individually, the results depicted in the plots are aggregated over the individual experiments. Moreover, we ran the QFL experiments in two different settings in each topology; results of both are shown in each plot. In one setting, the QPU capacity (i.e., number of qubits) varies for each client. As QPUs therefore run slightly different circuits, the number of features embedded also vary. To accommodate this, the images were resized for each client individually such that it fits the capacity of the respective client. Details on number of qubits and embedded features for each client are listed in Table \ref{tab:parameters}. In the second setting, we ran the QFL experiments in which each client has the same capacity; the capacity was chosen such that the entire image could be embedded. The motivation behind the first approach is the fact that in a potential quantum internet, quantum computers of different architectures, capacity, etc. might be connected; we evaluated such a setting on a small scale.

\begin{figure}
    \begin{subfigure}{0.5\textwidth}
        \includegraphics[scale=0.4]{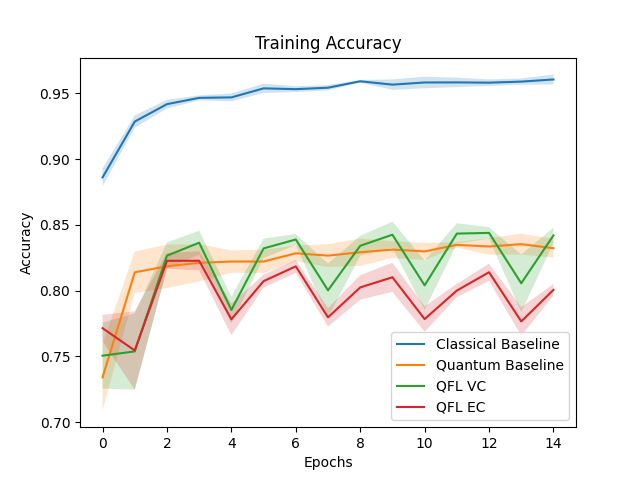}
        \caption{Training accuracy.}
        \label{fig:mcmd_star_train_accuracy}
    \end{subfigure}
    \begin{subfigure}{0.5\textwidth}
        \includegraphics[scale=0.4]{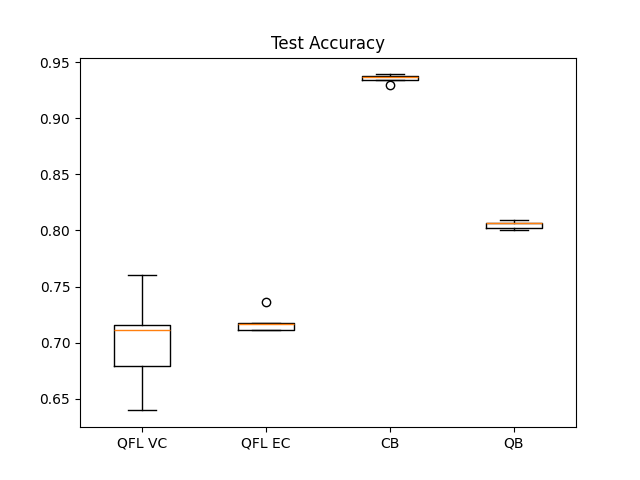}
        \caption{Test accuracy.}
        \label{fig:mcmd_star_test_accuracy}
    \end{subfigure}
    \caption{MCMD experiment results using the star topology.}
\end{figure}

Figures \ref{fig:mcmd_star_train_accuracy} and \ref{fig:mcmd_star_test_accuracy} show the training and test results using the star topology. While the QFL approach with varying QPU capacity delivers better training results than with equal capacity, its results on the test set are less stable.

\begin{figure}
    \begin{subfigure}{0.5\textwidth}
        \includegraphics[scale=0.4]{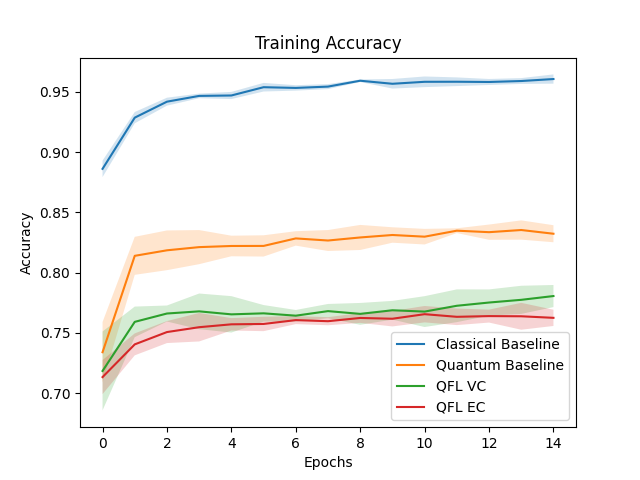}
        \caption{Training accuracy.}
        \label{fig:mcmd_ring_train_accuracy}
    \end{subfigure}
    \begin{subfigure}{0.5\textwidth}
        \includegraphics[scale=0.4]{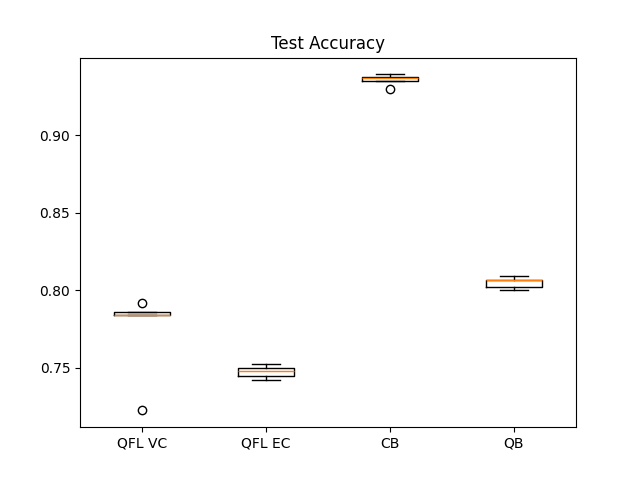}
        \caption{Test accuracy.}
        \label{fig:mcmd_ring_test_accuracy}
    \end{subfigure}
    \caption{MCMD results using the ring topology.}
\end{figure}

The results using the ring topology are shown in Figure \ref{fig:mcmd_ring_train_accuracy} and \ref{fig:mcmd_ring_test_accuracy}. In this approach, the quantum baseline delivers the best quantum results in both training and test accuracy. All approaches, including the classical baseline, converge relatively early though. 

Overall, the baselines deliver the best results in our experiments on image data, however, the QFL approaches produce acceptable results all the same. Though the differences in accuracy between both training approaches in QFL, i.e., ring and star, require further investigation. While the star approach seems less stable during training, the training accuracy nonetheless performs much better than the ring approach. However, on the test set the ring approach performs slightly better. Adjusting hyperparamters and the nature of weight exchange and aggregation are likely to affect the models performance, though optimizing for performance (i.e., accuracy) was not the objective of these experiments and is left for future work.

\section{Conclusion}\label{sec:conclusion}
Though the quantum internet promises diverse applications, it remains in nascent stages of development. Determining its exact nature, benefits, and optimal applications requires further comprehensive research. QFL is one such potential application, having its roots in classical machine learning, it can nonetheless be transferred to the field of quantum computing. And while QFL has been studied by the QML research community in recent years, far more research is required, especially when taking concrete ramifications of quantum communication into account as new challenges will inevitably arise through the incorporation of this novel communication medium.

In this paper, we presented and evaluated different scenarios of QFL that may arise on a potential quantum internet. We ran experiments using different network constraints that also differ how the clients are trained. Moreover, we ran these experiments in two settings, i.e., (1) QPUs with varying capacity and (2) equal capacity, where the former may be a more realistic setting in a large-scale quantum internet. Our results show that QFL is a viable alternative to regular training of quantum models, we furthermore show that the topology, i.e., the way models are trained influences the models performance. Note, however, we only simulated our experiments, moreover, weights were exchanged classically. Using quantum communication (e.g. teleportation) for the weight exchange should be investigated in future studies. Other relevant factors such as trainability and scalability should also be explored.

\begin{credits}
\subsubsection{\ackname}
This work is sponsored in part by the Bavarian Ministry of Economic Affairs, Regional Development and Energy as part of the 6GQT project (\url{https://6gqt.de})

\end{credits}

\bibliographystyle{splncs04}
\bibliography{bibliography}

\end{document}